\documentstyle[11pt]{article}

\begin{document}
\small

\title{Probabilistic Analysis of the 
Number Partitioning Problem}
\author{ F F Ferreira and J F Fontanari \\
Instituto de F\'{\i}sica de S\~ao Carlos\\
Universidade de S\~ao Paulo\\
Caixa Postal 369\\
13560-970 S\~ao Carlos SP\\
Brazil}
\date{}
\maketitle

\small

\bigskip


\centerline{\large{\bf Abstract}} \bigskip

Given a sequence of $N$ positive real numbers $\{ a_1,a_2,\ldots, a_N \}$,
the number partitioning problem consists of partitioning them into two sets
such that the absolute value of the difference of the sums of $a_j$ over the
two sets is minimized. 
In the case that the $a_j$'s are statistically independent random 
variables uniformly distributed in the unit interval, this NP-complete 
problem is equivalent to the problem
of finding the ground state of an infinite-range, random anti-ferromagnetic
Ising model. We employ the annealed approximation to derive 
analytical 
lower bounds to the average value
of the difference for the best constrained and unconstrained partitions in
the large $N$ limit. Furthermore, we calculate analytically the fraction of
metastable states, i.e. states that are  stable against all single spin 
flips, and found that it vanishes like  $N^{-3/2}$. 

\vspace{1cm}

{\bf Short Title:} number partitioning problem

\vspace{0.3cm} {\bf Physics Abstracts:} 87.10.+e - 64.60.Cn

\newpage



\section{Introduction}

The importance of the study of complex optimization problems 
which involve 
quenched, random, frustrated functions of many variables,
as well as the major role that statistical mechanics can 
play in that study,
have been pointed out by Anderson more than ten years ago \cite{Anderson}. 
Since then, standard statistical mechanics techniques have been applied
to the probabilistic analysis of  several classical combinatorial  
optimization  problems, such as
the graph partitioning problem \cite{AF}, 
the traveling salesman problem
\cite{MP,Cerf}, the knapsack problem \cite{Opper,Fonta,Jap}, 
and the satisfiability
problem \cite{Selman,Monasson,Brian}, to mention only a few. 
In fact the well-established statistical mechanics methods to 
characterize ground states (global minima) and metastable 
states (local minima) of spin glass models can be readily adapted to
the study of optimization problems \cite{MPV}.

In this paper we study the number partition problem (NPP) which 
is stated as follows.
Given a sequence of real numbers $\{ a_1,a_2,\ldots,a_N \}$, the 
NPP consists of partitioning them into two disjoint
sets ${\cal {A}}_1$ and ${\cal {A}}_2$ such that the difference
\begin{equation}
\mid \sum_{a_j \in {\cal {A}}_1 } a_j 
- \sum_{a_j \in {\cal {A}}_2 } a_j \mid
\end{equation}
is minimized. Alternatively, we can search for the Ising spin
configurations ${\bf s} =
\left ( s_1,\ldots,s_N \right ) $ that minimize the 
energy or cost function 
\begin{equation}  \label{E_1}
E \left ( {\bf s} \right ) = ~ \mid \sum_{j=1}^N a_j s_j \mid,
\end{equation}
where $s_j = 1$ if $a_j \in {\cal {A}}_1 $ and $s_j = -1$ if 
$a_j \in {\cal {A}}_2 $. We can consider also the problem 
of constrained partitions, in which the difference between 
the cardinalities of sets 
${\cal {A}}_1$ and $ {\cal {A}}_2$ is fixed, i.e., 
\begin{equation}  \label{m}
m = \frac{1}{N} ~\sum_{j=1}^N s_j  .
\end{equation}
The NPP may be viewed as the practical problem of finding the 
fairest way to
partition a set of $N$ objects $j=1,2, \ldots,N$, each of which 
of value $a_j$, between two persons. Despite its simplicity, 
the NPP was shown to 
belong to the NP-complete class, which basically means that there 
is no known deterministic algorithm guaranteed to solve all instances 
of this problem within a polynomial time bound \cite{GJ}.
The fact that the NPP is frustrated can easily be
understood by squaring equation (\ref{E_1}), so that the problem of
minimizing $E$ becomes then the one of finding the ground state of the
infinite-range, random anti-ferromagnetic Ising Hamiltonian \cite{Fu}
\begin{equation}\label{H}
{\cal{H}} = \frac{1}{2} \sum_{i} \sum_{j>i} a_i a_j s_i s_j .
\end{equation}
Thus we note that the problem of finding the ground 
state of (\ref{H}) is  NP-complete.

Although zero-cost solutions of the NPP may be of some value to 
cryptography \cite{Shamir}, the interest in this problem stems 
mainly from the remarkable failure of the stochastic heuristic 
simulated annealing \cite{KGV,Pablo}
to find good solutions to it, as compared
with the solutions found by deterministic heuristics \cite{JAMS}. 
In fact, the reason
for that failure is that the usual strategy of exploring the space 
of configurations $\{{\bf s}\}$ through single spin flips leads 
to changes of energy that are typically of order $1/N$, while a 
theoretical analysis 
indicates that the global minimum energy is of
order $\sqrt{N}~2^{-N}$ for unconstrained partitions \cite{KKLO}. 
It is interesting to note that a very simple deterministic 
heuristic, the differencing method
of Karmakar and Karp \cite{KK}, can find with high probability 
solutions whose energies are of
order $1/N^{\alpha \log N}$ for some $\alpha >0$. 
More recently, it has been
shown that the performance of simulated annealing can be 
greatly improved and even surpass that of the differencing method  
by employing different representations for the problem \cite{Ruml}.

In this work we employ the annealed approximation \cite{TV,VM}
to derive rigorous lower bounds to the average value of the 
difference or energy for the best constrained and unconstrained 
partitions. For constrained partitions, we show that
the average optimal energy
is extensive for  $ m > \sqrt{2}-1$ and 
we calculate it exactly in this regime using the 
self-averaging property of
the free energy density. The theoretical
predictions are compared with numerical estimates for the
optimal energy obtained  through the
exhaustive search of the configuration space for $N \leq 24$.
Furthermore, we calculate analytically the average 
number of minima in the 1-swap neighborhood and
estimate their typical energy. A minimum in the 1-swap 
neighborhood is a state that has lower energy than
all the $N$ states that differ from it  by a single spin only 
\cite{JAMS}.
Similarly to previous studies of 
the NPP \cite{JAMS,KKLO,Ruml}, 
we will consider  the case where the $a_j$'s are statistically 
independent random variables uniformly distributed in the unit interval.

The remainder of this paper is organized as follows. In section 2
we describe the annealed approximation and calculate the lower
bounds to the average value of the optimal energy.  
In section 3 we present the calculation of the average number of 
local minima in the 1-swap neighborhood. In section 4 we
discuss our main results and present
some concluding remarks. In particular, we 
compare our approach with  other theoretical studies of 
the NPP \cite{Fu,KKLO}. In the appendix we present the details
of the self-averaging calculation of the average optimal
energy in the regime where this quantity is extensive.


\section{Annealed approximation}


In the canonical ensemble formalism of the statistical mechanics
the average value of the optimal energy for
constrained partitions is given by
\begin{equation}\label{E_m}
\bar{E}_m =  \lim_{T \rightarrow 0}  F_m (T) = -
\lim_{T \rightarrow 0} T ~ \left \langle \ln Z_m \right \rangle ,
\end{equation}
where $F_m (T)$ is the average free energy, and
 $Z_m (T)$ is the partition function
\begin{equation}\label{Z_m}
Z_m (T) = \sum_{\{ {\bf s} \} } \delta \left ( N m,  \sum_j s_j 
\right ) \exp \left [ - \frac{E \left ( {\bf s} \right )}{T} \right ]
\end{equation}
with $m= -1, -1 +2/N, \ldots, 1-2/N, 1$. 
Here
the summation is over the $2^N$ states ${\bf s}$, 
$\delta (k,l)$ is the Kronecker delta and  $T$ 
is the temperature. The notation $\langle \ldots \rangle$
stands for the average over the  random variables $a_i$. 
The limit $T \rightarrow 0$ in equation (\ref{E_m})
ensures that only the states that minimize $E \left ( {\bf s} \right )$
will contribute to $Z_m$.

Since the average entropy $S_m (T) = - dF_m/dT $
of a system of Ising spins is positive at all temperatures,
$F_m$ must be a decreasing function of $T$, so that
$\bar{E}_m = F(0) \geq F(T)$ for all $T$. 
Defining the annealed free energy by
\begin{equation}\label{F_m^a}
F_m^a (T) = - T ~\ln ~ \left \langle Z_m (T) \right \rangle ,
\end{equation}
and using Jensen's inequality \cite{Feller}, 
$\ln \langle Z_m \rangle \geq \langle \ln Z_m \rangle$,
yield the following
inequalities
\begin{equation}
F_m^a (T) \leq F_m (T) \leq \bar{E}_m . 
\end{equation}
Thus,  the annealed
free energy calculated at any $T$ provides a rigorous lower 
bound to $\bar{E}_m$ \cite{TV,VM}. 
Clearly,
the tightest bound is given by $\bar{E}_m^a = F_m^a (T_m^*)$ 
where $T_m^*$ is
the temperature that maximizes $F_m^a (T)$, i.e.  
\begin{equation}\label{derivative}
\frac{d F_m^a}{dT}  \mid_{T_m^*} =0 .
\end{equation}
This procedure is very useful because, in general,
the annealed free energy 
is much easier to evaluate than the quenched one.

We now proceed with the explicit evaluation of the annealed free energy.
Using the integral representations of the Dirac and Kronecker delta 
functions we write 
\begin{eqnarray}\label{Z_m_1}
\langle Z_m (T) \rangle & = & \int_{-\infty}^\infty 
 \int_{-\infty}^\infty
 \frac{dx d\tilde{x}}{%
2 \pi} \, \mbox{e}^{i x \tilde{x} - \mid x \mid /T } \, 
\int_{-\pi}^\pi \frac{d\tilde{m}}{2 \pi} \, \mbox{e}^{i N m 
\tilde{m}} \,   \nonumber \\
& & \prod_j \int_0^1 da_j \sum_{s_j = \pm 1} \exp 
\left [ - i s_j \left (
a_j \tilde{x} + \tilde{m} \right ) \right ] .
\end{eqnarray}
The
integrals over $x$ and $a_j$, as well as the
summation over $s_j$, can easily be performed yielding
\begin{eqnarray} 
\langle Z_m (T) \rangle & = & \int_{-\infty}^{\infty} 
\frac{d\tilde{x}}{2 \pi} \, \frac{ 2 T}{ 1 + \left ( T \tilde{x}\right )^2}
 \,
\left [ \frac{\sin \left (\tilde{x}/2 \right )}{\tilde{x}/2} \right ]^N
\nonumber \\
& &
\int_{-\pi}^{\pi}  \frac{d\tilde{m}}{2 \pi} \mbox{e}^{i N m \tilde{m}} \,
\left [ \mbox{e}^{i \tilde{m} + i \tilde{x}/2} + 
 \mbox{e}^{-i \tilde{m} - i \tilde{x}/2} \right ]^N .
\end{eqnarray}
Using the binomial theorem, 
the integral over $\tilde{m}$ can be readily carried out. The final 
result is simply
\begin{equation}\label{Z_m_2}
 \langle Z_m (T) \rangle  =  \left ( \! \! \begin{array}{c} N \\ n 
\end{array} \! \! \right ) \, \int_{-\infty}^{\infty} 
\frac{dy}{\pi} \, \frac{ 2 T}{ 1 + \left ( 2 T y \right )^2} \,
\mbox{e}^{N G_m (y)} 
\end{equation}
where 
\begin{equation}
n = N ~ \frac{1-m}{2} ,
\end{equation}
\begin{equation}
G_m (y) = i m y + \ln \left ( \frac{\sin y}{y} \right ) ,
\end{equation}
and we have made the change of variable $y = \tilde{x}/2$. 
In the limit of large $N$, the integral over $y$ can be 
evaluated using 
the saddle-point method \cite{Daniels}. Since $\mid m \mid \leq 1$,
the saddle-point is the imaginary $y_s = i \zeta$, where $\zeta$ 
is the real
solution of the equation
\begin{equation}
m - \mbox{coth} ~\zeta + \frac{1}{\zeta} = 0 .
\end{equation}
Hence, the function $G_m (y_s) = G_m$, where
\begin{equation}
G_m  = - m \zeta + \ln \frac{\sinh \zeta}{\zeta}  ,
\end{equation}
is real. Finally, using Stirling's formula for the binomial coefficient
we rewrite equation (\ref{Z_m_2}) in the limit of large $N$ as
\begin{equation}\label{ZZ}
\langle Z_m (T) \rangle = \frac{2}{\pi N} 
\sqrt{ \frac{1}{(1-m^2) \mid G_m^{''} \mid }}~
\frac{2 T}{1 - \left ( 2 T \zeta \right )^2}
 ~ \mbox{e}^{N g_m}
\end{equation}
where
\begin{equation}
g_m = G_m  - \frac{1+m}{2} \ln \frac{1+m}{2} 
- \frac{1-m}{2} \ln \frac{1-m}{2} 
\end{equation}
and
\begin{equation}
G_m^{''} = -1 + m^2 + \frac{2m}{\zeta} .
\end{equation}
At this stage we can readily calculate the temperature $T_m^*$ that
maximizes the annealed free energy. In fact, equation 
(\ref{derivative}) is written as
\begin{equation}\label{T^*_1}
 \ln \left \langle Z_m \left ( T^*_m \right ) \right \rangle +
\frac{ 1 + \left ( 2 T^*_m \zeta \right )^2}
{1 - \left ( 2 T^*_m \zeta \right )^2} = 0 .
\end{equation}

We consider first the regime where
$\left \langle Z_m \left ( T^*_m \right ) \right \rangle$
is of order $1$. In this case, equation (\ref{ZZ})
implies that $T^*_m$ is vanishingly
small,  so that equation (\ref{T^*_1}) reduces to
$ \left \langle Z_m \left ( T^*_m \right ) \right \rangle
= \mbox{e}^{-1}$. Inserting this result into 
equation (\ref{F_m^a}) yields $\bar{E}_m^a = T_m^*$. Hence,
\begin{equation}\label{lb_m}
\bar{E}_m^a = \frac{\pi N}{4} \sqrt{ (1-m^2) \mid G^{''}_m \mid } 
~\mbox{e}^{-1 - N g_m}
\end{equation}
which is consistent with the assumption that $T^*_m$ is small 
for large $N$, provided that $g_m > 0$. Since $g_m$ 
decreases monotonically with $m$, from $g_0 = \ln 2$ to $g_1 = -\infty$,
this assumption breaks down for $\mid m \mid > 0.560$ where 
$g_m$ is negative. Henceforth we will assume that $m \geq 0$.

It is instructive to consider in detail  the case of even partitions
($m = 0$). In this case we find $\zeta=0$, $g_0 = \ln 2$, and
$G_0^{''} = -1/3$ so that
\begin{equation}
\langle Z_0 (T) \rangle = 2^N T \frac{4 \sqrt{3}}{\pi N} 
\end{equation}
and
\begin{equation}\label{E_0^a}
\bar{E}^a_0 = 2^{-N} \frac{\pi N}{4 \,  \mbox{e} \sqrt{3}} 
\approx 0.167 ~2^{-N} N .
\end{equation}
In figures $ 1(a)$ and $1(b)$ we present the results of numerical
experiments to estimate the energy of the 
global minima for even partitions
through the
exhaustive search in  the configuration  space for $N \leq 24$. 
In all experiments discussed in this work, the symbols
represent the averages over $10^4$ realizations of the set $\{a_j \}$.
The error bars are calculated by measuring the standard deviation
of the average optimal energies obtained in $25$ experiments, each one 
involving the average of  $400$ realizations of the set $\{a_j \}$.
In these experiments we  focus on the 
$N$ dependence of the average optimal energy
$\bar{E}_m = \langle E_m \rangle $, and
of the ratio $r_m = \sqrt{\sigma_m^2}/\bar{E}_m$ 
where $\sigma_m^2 = \langle E_m^2 \rangle -\langle E_m \rangle^2$
is the variance of the random variable $E_m$. 
In figure $1(a)$ we show $\bar{E}_0$ as a function of $N$. The 
straight line shown in this figure yields the  fitting
$\bar{E}_0 = 0.80 ~2^{-N} N$.
Hence,  although the annealed bound $\bar{E}_0^a$ gives
the correct scaling with $N$, it is about five times smaller than
our  numerical estimate for $\bar{E}_0$.
In figure $1(b)$ we show the ratio $r_0$
as a function of $N$. Interestingly, this ratio tends to 1 for 
large $N$ indicating then that the optimal energy $E_0$ is not
self-averaging.

In the regime where 
$\left \langle Z_m \left ( T^*_m \right ) \right \rangle$
is of order $\mbox{e}^N$ we find 
$\ln \left \langle Z_m \left ( T^*_m \right ) \right \rangle
\approx N g_m$ and $T_m^* \approx 1/2 \zeta$ 
so that
\begin{equation}\label{EG}
\bar{E}_m^a = - N ~\frac{g_m}{2 \zeta} .
\end{equation}
Of course, this solution is valid only for $ m > 0.560$
where $g_m$ is negative. We note that (\ref{EG}) gives
a very poor lower bond to $\bar{E}_m$. In particular, for $m = 1$ we have 
$\bar{E}_1 = N/2$ while the annealed bound yields $\bar{E}_1^ a = 0$.
Fortunately, in the regime of extensive  $E_m $   we can 
use the self-averaging property of the free energy density
to calculate $\bar{E}_m$ exactly for large $N$ (see Appendix).
The final result is simply
\begin{equation}\label{self}
\bar{E}_m = \frac{N}{2} \left [ \frac{ \left ( 1 + m \right )^2}{2} - 1 
\right ] ,
\end{equation}
which is valid for $ m \geq \sqrt{2} - 1 \approx 0.414$.
Thus the annealed lower bound is also very poor in the region
$0.414 < m < 0.560$  since  in
this region $\bar{E}_m^a$ decreases exponentially with 
$N$, while $\bar{E}_m$ actually increases linearly with $N$.

To better appreciate the qualitative differences between the 
regimes of distinct scalings with $N$,  we present 
in figure $2$ the numerical estimates for $\bar{E}_m$ 
as a function of $m$ for $N=24$.
The existence of two different regimes of scaling with $N$,
as well as the very good agreement with the 
theoretical predictions for
$m > 0.414$, are apparent in this figure.
A noteworthy feature
of our numerical estimate for $\bar{E}_m$ shown in the inset
is that, in contrast
to the annealed lower bound (\ref{lb_m}), the even
partitions ($m=0$) do not give the lowest energy. We have
verified that this
result holds for smaller values of $N$ as well.
Furthermore,  there seems to occur a rather
abrupt transition at $m \approx 0.25$ as indicated by the large 
error bar and for the change of almost three orders
of magnitude in $\bar{E}_m$. Although it would be very interesting to 
study these results more carefully for larger $N$,
we are not aware of any efficient heuristic to solve the NPP for
constrained partitions.
In particular, we note that the differencing method \cite{KK}
applies only to unconstrained partitions.

We turn now to the analysis of unconstrained partitions.
The average partition function in this case is  given by
\begin{eqnarray}  \label{Z_u}
\langle Z_u (T) \rangle  & = &  \sum_m ~ \langle Z_m (T) \rangle 
\nonumber \\
& =  & 2^N \int_{-\infty}^{\infty} 
\frac{dy}{\pi} \, \frac{ 2 T}{ 1 + \left ( 2 T y \right )^2} \,
\mbox{e}^{N G_u (y)} ,
\end{eqnarray}
where
\begin{equation}
G_u (y) = \ln \left [ \frac{\sin \left (2y\right)}{2y} \right ] .
\end{equation}
As before, in the limit of large $N$
the integral over $y$ can be carried out via a saddle-point 
integration. Since the saddle-point is $y_s = 0$, the final result
is simply
\begin{equation}
\langle Z_u (T) \rangle = 2^N T \sqrt{\frac{6}{\pi N}} ,
\end{equation}
which yields 
\begin{equation}\label{E_u^a}
\bar{E}_u^a =  2^{-N} \sqrt{\frac{\pi N}{6 \mbox{e}^2}}
\approx 0.266 ~ 2^{-N} \sqrt{N} .
\end{equation}
It is interesting to compare this result with the average
energy  of a randomly chosen configuration ${\bf s}$.
This quantity, which is defined by
\begin{equation}
\bar{E}_r = 2^{-N} ~\prod_i \int_0^1 da_i ~ \sum_{s_i =\pm 1}
\mid \sum_i a_i s_i \mid ,
\end{equation}
is easily calculated and yields $\bar{E}_r = \sqrt{2N/3 \pi}$ 
for large $N$. 
Moreover, comparing equations (\ref{E_0^a}) and (\ref{E_u^a}) 
we find that
the lower bound for average optimal energy of even partitions ($m=0$),
which minimizes $E_m^a$, is larger than
that of unconstrained partitions by a factor
$ N^{1/2}$. The fact that these quantities do not coincide
indicates that,  for unconstrained partitions, $m$ is not a 
self-averaging quantity, even in the large $N$ limit, i.e.
the values of $m$ associated to the best unconstrained partitions
depend on the specific realization of the set of random variables
$\{a_j \}$. In figure $3(a)$ we present the numerical estimate for
the average optimal energy $\bar{E}_u = \langle E_u \rangle$ 
obtained through the exhaustive search for $N \leq 24$.
The data are very
well fitted by $\bar{E}_u  = 1.12  ~2^{-N} \sqrt{N} $. 
In figure $3(b)$ we show the ratio $r_u = \sqrt{\sigma_u^2}/\bar{E}_u$
as a function of $N$. As before, the
finding that this ratio tends to 1 for increasing $N$
indicates that  $E_u$ is not  self-averaging.


\section{Average number of local minima}


As mentioned before, a minimum in the 1-swap 
neighborhood is a state that has lower energy than
all the $N$ states that differ from it  by a single spin only 
\cite{JAMS}. In the statistical mechanics context, these states
are usually termed metastable states \cite{Tanaka}. 
The following analysis will
be restricted to unconstrained partitions only, since for constrained
partitions we would have to consider
the simultaneous flip of two spins in order
to satisfy the cardinality constraint.
The average number local minima with energy $E = \mid t \mid $  
is defined by
\begin{equation}
\langle {\cal{M}} \left ( t \right )  \rangle = 
\left \langle \sum_{\{ {\bf s} \} }
\delta \left ( t - \sum_j s_j a_j \right )
\prod_i \Theta \left ( \mid t - 2 s_i a_i \mid
- \mid t \mid \right ) \right \rangle
\end{equation}
where $\delta (x) $ is the Dirac delta function and
$\Theta (x) = 1$ if $x \geq 0$ and $0$ otherwise. 
As the calculation is straightforward we will only sketch it in
the sequel.  Using the integral representation of the
delta function we obtain 
\begin{equation}
\langle {\cal{M}} \left ( t \right ) \rangle = 
\int_{-\infty}^\infty \frac{d \tilde{t}}{2 \pi} 
~\mbox{e}^{i t \tilde{t}} \prod_j  \sum_{s_j= \pm 1} 
\int_0^1 da_j ~\mbox{e}^{-i \tilde{t} s_j a_j} ~
\Theta \left ( \mid t - 2 s_j a_j \mid - \mid t \mid \right ) .
\end{equation}
Hence
the integral over $a_j$ and the summation over
$s_j$ can readily be performed, yielding
\begin{equation}\label{El1}
\langle {\cal{M}} \left ( t \right ) \rangle  =  
\int_{-\infty}^{\infty} 
\frac{d \tilde{t}}{2\pi}  ~\mbox{e}^{i t \tilde{t} }
\left ( \frac{\mbox{e}^{-i t \tilde{t} } - \mbox{e}^{i \tilde{t} }
+ \mbox{e}^{i \tilde{t} } - 1}{i \tilde{t}} \right )^N 
~~~~~\mbox{if}~~ E = \mid t \mid ~ < 1 ,
\end{equation}
and 
\begin{equation}\label{Eg1}
\langle {\cal{M}} \left ( t \right ) \rangle  =  
\int_{-\infty}^{\infty} 
\frac{d \tilde{t}}{2 \pi}  ~\mbox{e}^{i t \tilde{t} }
\left ( \frac{\mbox{e}^{i \tilde{t}}- 1}{i \tilde{t}} \right )^N 
= 0
~~~~~\mbox{if}~~ E = \mid t \mid ~ \geq 1 ,
\end{equation}
where we have used the interesting result  that
the integral in equation (\ref{Eg1})
vanishes for all $N$ \cite{tabela}. Thus, there are no local minima 
with 
$E \geq 1$. As usual, for large $N$ the integral in equation (\ref{El1})
can evaluated  via a saddle-point integration. 
The final result is
\begin{equation}
\langle {\cal{M}} \left ( t \right ) \rangle  =
\sqrt{ \frac{1}{2 \pi N \mid H''(\xi) \mid } } 
~ \mbox{e}^{ N H (\xi) }
\end{equation}
where
\begin{equation}
H ( \xi ) = \ln 2 + \ln \left [ \frac{\sinh \xi}{\xi} -
\mbox{e}^{- t \xi/2}~ \frac{\sinh \left ( t \xi/2 \right )}{\xi}
\right ] ,
\end{equation}
and $ H''(\xi) = - d^2 H (\xi)/d \xi^2$. Here,
$\xi$ is the solution of
\begin{equation}
\frac{2}{\xi} - \frac{2 \cosh \xi - \mbox{e}^{- t \xi} ~t }
{\sinh \xi - \mbox{e}^{- t \xi/2} ~\sinh \left ( t \xi/2 \right ) }
= 0 .
\end{equation}
The function $H (\xi)$ is a monotonically decreasing function
of $E = \mid t \mid$. In particular, it decreases from $\ln 2$ at  
$E = 0$ ($\xi = 0$)
to $- \infty$ at $E = 1$ ($\xi = - \infty$). 
It vanishes
at $E  \approx 0.851$, so the average number of local minima with
energy larger than  that value decreases
exponentially with $N$.

A more interesting quantity is
the average number of local minima regardless their energy
values, which is defined by
\begin{equation}\label{total_1}
\langle {\cal{M}}  \rangle = \int_{-\infty}^\infty dt ~
\langle {\cal{M}} \left ( t \right ) \rangle .
\end{equation}
From the above discussion,
it is clear  that only the close neighborhood of 
$t=0$ contributes to this integral, so we can expand the integrand
of (\ref{El1}) in powers of $t$ and $\tilde{t}$ and keep the lowest
order terms only. The final result is 
\begin{equation}\label{total_2}
\langle {\cal{M}} \rangle = \sqrt{\frac{24}{\pi}}~\frac{2^N}{N^{3/2}}
  \approx 2.764~\frac{2^N}{N^{3/2}} .
\end{equation}
It is interesting to estimate the dependence on
$N$ of the typical energy of a local minimum. This quantity,
denoted by $E_t$, is defined by
\begin{equation}
E_t =  \langle \frac{ \int dt~ t ~ {\cal M} (t) }
{ \int dt~  {\cal M} (t)}  \rangle ,
\end{equation}
which, in the annealed approximation framework \cite{Tanaka},
is approximated by
\begin{equation}\label{E_t_a}
E_t \approx   \frac{\int dt~ t ~ \langle {\cal M} (t) \rangle }
{\langle {\cal M}   \rangle} .
\end{equation}
The procedure to evaluate (\ref{E_t_a}) is identical to that used
in the evaluation of (\ref{total_1}) and yields 
\begin{equation}\label{approx}
E_t \approx   \frac{2}{N} .
\end{equation}
We note that while equation (\ref{total_2}) gives the exact 
leading order term of the average number of local minima, equation
(\ref{approx}) is an uncontrolled estimate for the energy of a
typical minimum. These quantities can be easily estimated 
numerically: for each value of $N$, ranging from $100$ to $3000$,
we generate $10^5$ random states ${\bf s}$ and count
the fraction of them that are local minima and measure their energies.
We find that the numerical data are very well fitted by the equations 
$ \langle {\cal M} \rangle 
\approx (2.81 \pm 0.02)~2^N/N^{3/2}$
and $E_t \approx (1.76 \pm 0.04)/N$, which are in quite good agreement
with the theoretical predictions.


\section{Conclusion}


To appreciate some of the drastic features of the energy landscape
associated to the NPP or, equivalently, to the random anti-ferromagnetic
Ising model defined by the Hamiltonian (\ref{H}), 
we compare our results with those of the
SK model, which is
defined by the Hamiltonian \cite{SK}
\begin{equation}
{\cal H} = - \sum_i \sum_{j > i}J_{ij} s_i s_j ,
\end{equation}
where the couplings $J_{ij}$ are Gaussian statistically independent 
random variables of zero mean and  variance $1/N$. In this model the 
annealed 
lower bound for the
ground state energy is $E^a = - 0.833 N$ \cite{TV} and the number
of metastable states increases as $\mbox{e}^{0.199 N}$ \cite{Bray}.
Hence, in the NPP  there are much more local minima  
and the global minima are much deeper  than in the SK model. These
findings may explain the failure of local 
search techniques 
to produce good solutions  to the NPP.

Some comments regarding the comparison of our approach with that 
of Karmakar  {\em et al.} \cite {KKLO} are in order. 
Those authors have  derived bounds on the
probability of occurrence of the event
$ {\cal {N}} (E) = 0 $, where ${\cal {N}} (E)$ stands for
the number of states ${\bf s}$ with energy smaller
than $E$. 
Interestingly, these bounds are related to the first two
moments of ${\cal {N}}$: 
\begin{equation}
1 - \langle {\cal {N}} \rangle 
\leq Pr \{ {\cal {N}} = 0 \} \leq \frac{
\langle {\cal {N}}^2 \rangle - \langle {\cal {N}} \rangle^2 }{\langle 
{\cal {N}}^2 \rangle} .
\end{equation}
The first inequality follows
trivially from the fact that ${\cal {N}} \geq 0$, 
while the second is an improvement of Chebyshev's inequality. 
Only unconstrained and even partitions ($m=0$) were considered.
However, as acknowledged by Karmakar 
{\em et al.} \cite{KKLO}, these bounds give no information 
on the average value
of the difference for the best partition, except perhaps for 
its scaling with $N$.
Also, we should mention that Fu \cite{Fu} has actually carried out a
replica analysis of the NPP for the case of even partitions $m=0$.
However, since in that analysis it is assumed that
$E_0$ is extensive, it misses the low-temperature phase 
completely.

%

\bigskip

 {\bf Acknowledgments}~~JFF thanks Pablo Moscato
for useful conversations. This work was supported in part by 
Conselho Nacional
de Desenvolvimento Cient\'{\i}fico e Tecnol\'ogico (CNPq).


\newpage

\section*{Appendix}

In this appendix we calculate exactly the average optimal energy
in the regime where $E_m$ scales linearly with $N$. 
Similarly to equation (\ref{Z_m_1}) we write the partition function
defined in (\ref{Z_m}) as
\begin{eqnarray}
 Z_m (T)  & = & \int_{-\infty}^\infty 
 \int_{-\infty}^\infty
 \frac{dx d\tilde{x}}{%
2 \pi} \, \mbox{e}^{i x \tilde{x} - \beta \mid x \mid } \, 
\int_{-\pi}^\pi \frac{d\tilde{m}}{2 \pi} \, \mbox{e}^{i N m 
\tilde{m}} \,   \nonumber \\
& & \prod_j  \sum_{s_j = \pm 1} \exp 
\left [ - i s_j \left (
a_j \tilde{x} + \tilde{m} \right ) \right ] ,
\end{eqnarray}
where $\beta = 1/T$ is the inverse temperature. As in the
annealed approximation,
the summation over $s_j$ can easily be performed,
yielding
\begin{eqnarray}
 Z_m (T) &  = &  N \beta^2
\int_{-\infty}^\infty dx \int_{-i \infty}^{i \infty} 
\frac{d\tilde{x}}{2 \pi i} \, \int_{-i \pi/\beta}^{i \pi/\beta}
\frac{d\tilde{m}}{2 \pi i} 
\exp \left [- N \beta \left (x \tilde{x} + \mid x \mid + m  \tilde{m}
\right) \right ] \nonumber \\
& & \times
 \exp \left [ N \int_0^1 da ~ 
\ln 2 \cosh \beta \left ( \tilde{x} a + \tilde{m} \right ) 
\right ] ,
\end{eqnarray}
where we have used the self-averaging property
\begin{equation}
\frac{1}{N} \sum_j 
\ln 2 \cosh \beta \left( \tilde{x} a_j + \tilde{m} \right )
= \int_0^1 da 
\ln 2 \cosh \beta \left( \tilde{x} a + \tilde{m} \right ) ,
\end{equation}
which is exact for $N \rightarrow \infty$. In this limit we
can carry out  the
integrals using the saddle-point method, and so we obtain the following
equation for the average free-energy density 
$\bar{f}_m = \bar{F}_m/N$:
\begin{equation}\label{f_m}
\bar{f}_m =  x \tilde{x} + \mid x \mid + m \tilde{m} -
\frac{1}{\beta} ~\int_0^1 da 
\ln 2 \cosh \beta \left( \tilde{x} a + \tilde{m} \right ) .
\end{equation}
In the zero-temperature limit ($\beta \rightarrow
\infty$), the saddle-point equations yield $\tilde{x} = -1$,
$\tilde{m} = (1+m)/2$ and
\begin{equation}
x = \frac{\left ( 1 + m \right )^2}{4} -  \frac{1}{2} ,
\end{equation}
where we have assumed $x \geq 0$ and $m \geq 0$. 
The average optimal energy
is obtained by taking 
the zero-temperature limit in equation (\ref{f_m}) which yields
$\bar{f}_m \rightarrow \bar{E}_m /N = \mid x \mid$.


\newpage 

\section*{Figure captions}

\bigskip

\parindent=0pt 

{\bf Fig.\ 1(a)} Average optimal energy obtained
through exhaustive search as a function of $N$ for even
partitions ($m=0$).

\bigskip

{\bf Fig.\ 1(b)} Ratio between the standard deviation
and the average value of the optimal energy obtained
through exhaustive search  
as a function of $N$ for even partitions ($m=0$).

\bigskip

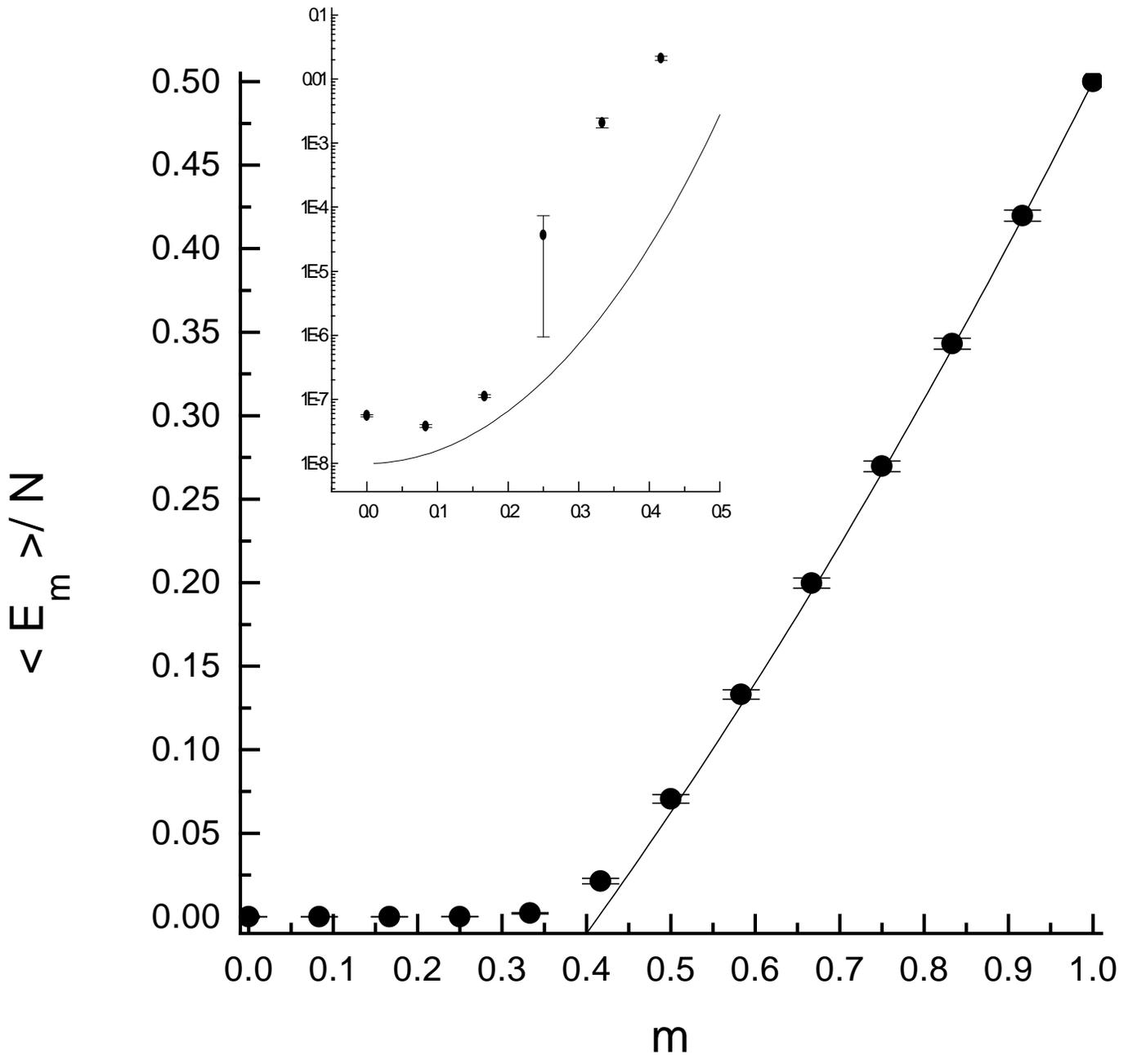
{\bf Fig.\ 2} Average optimal energy obtained
through exhaustive search as a function of $m$ for $N=24$.
The full curve in the principal graph
is the theoretical estimate (\ref{self}) while
the one in the inset is the annealed lower bound (\ref{lb_m}).

\bigskip

{\bf Fig.\ 3(a)} Average optimal energy obtained
through exhaustive search as a function of $N^{1/2}$ 
for unconstrained partitions.

\bigskip

{\bf Fig.\ 3(b)} Same as figure $1(b)$ but for unconstrained
partitions.


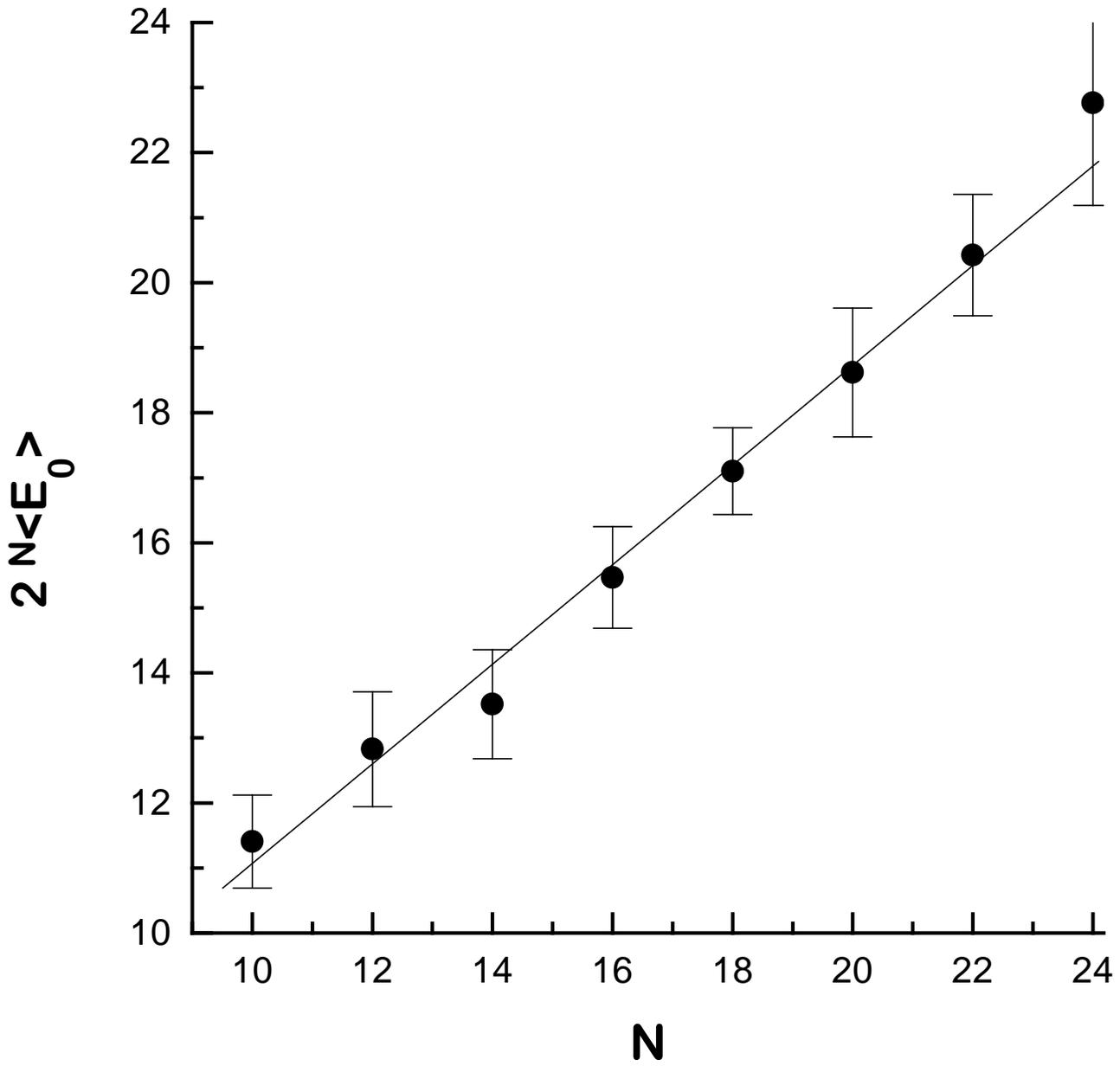
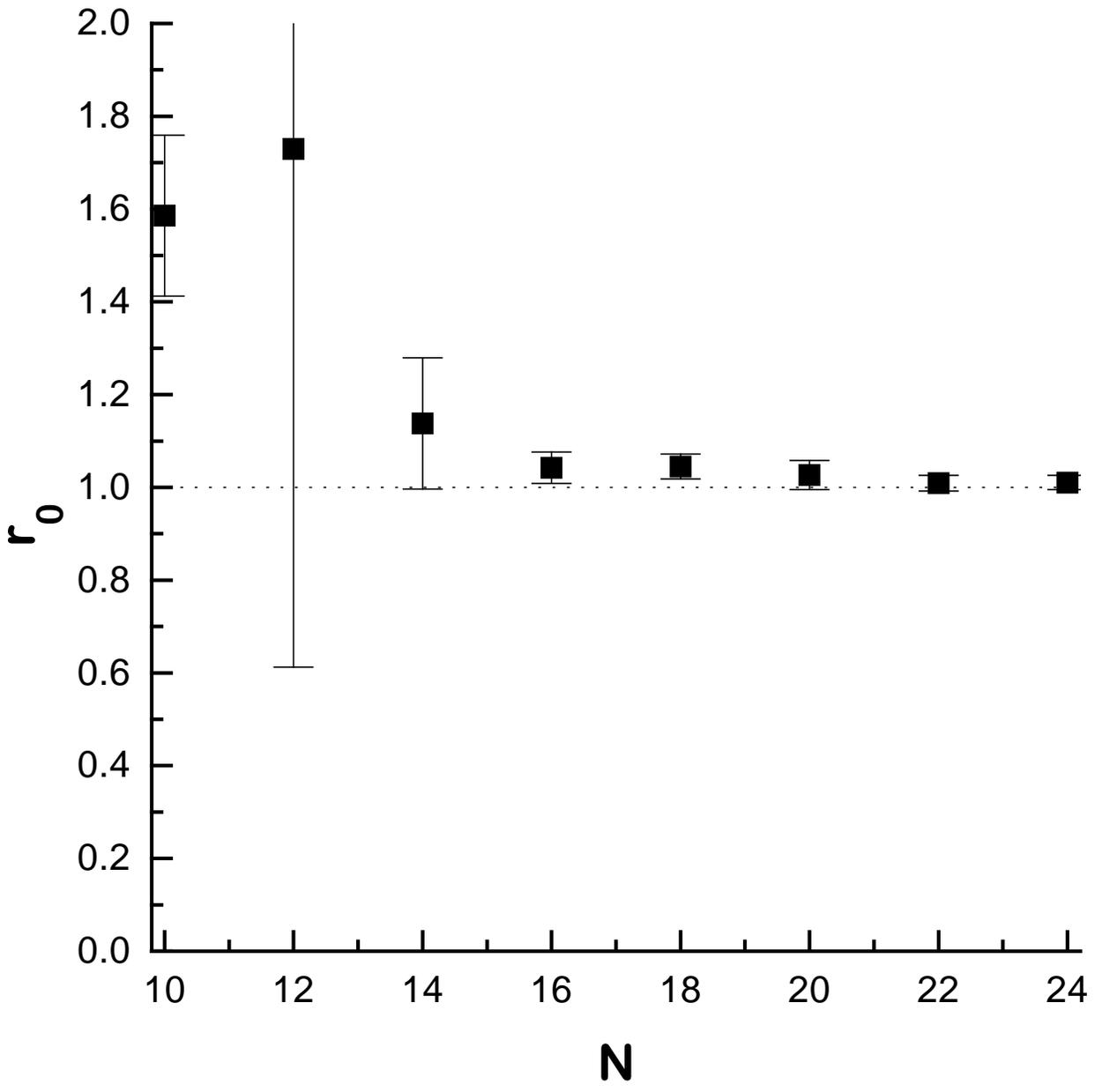
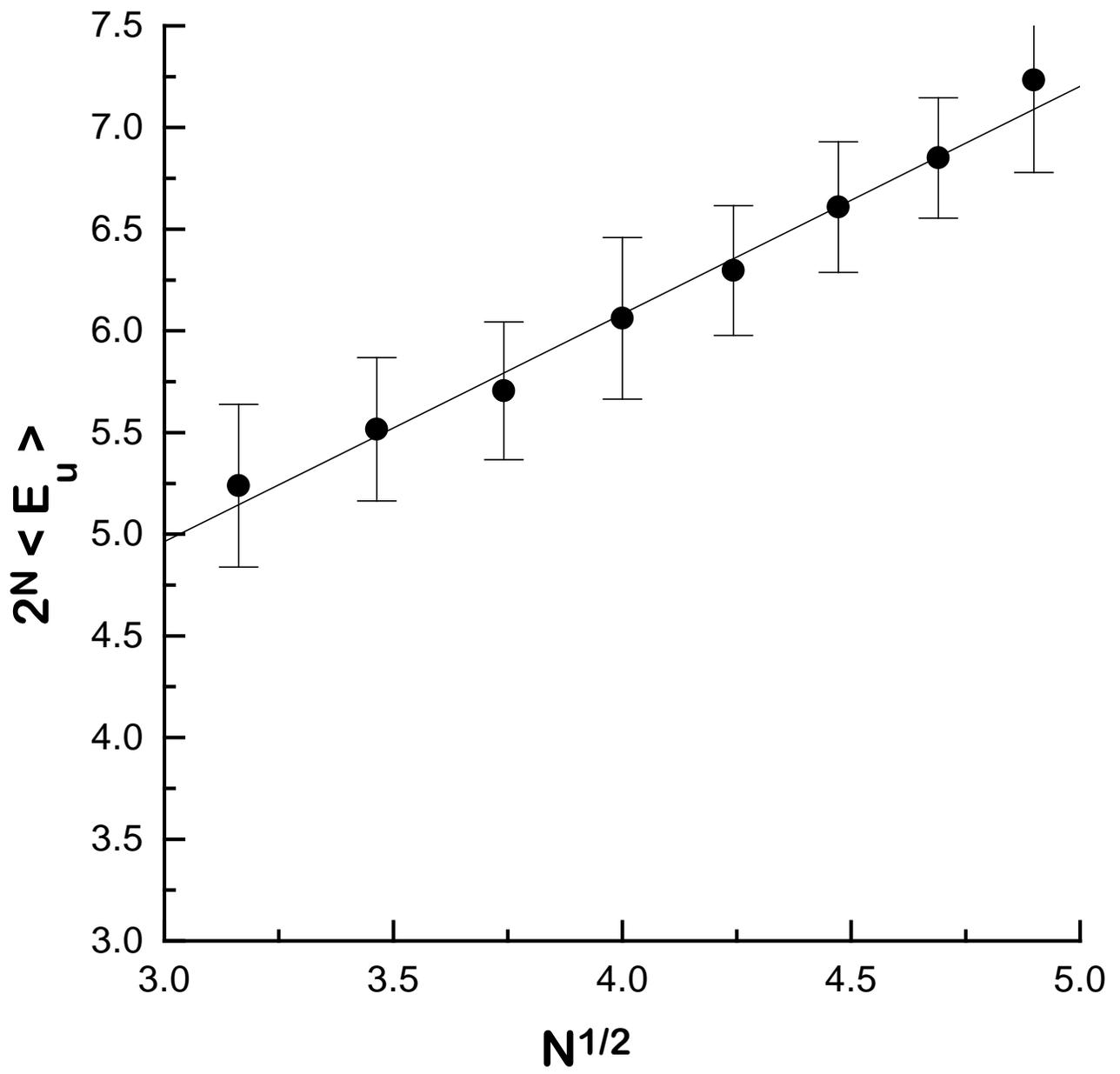
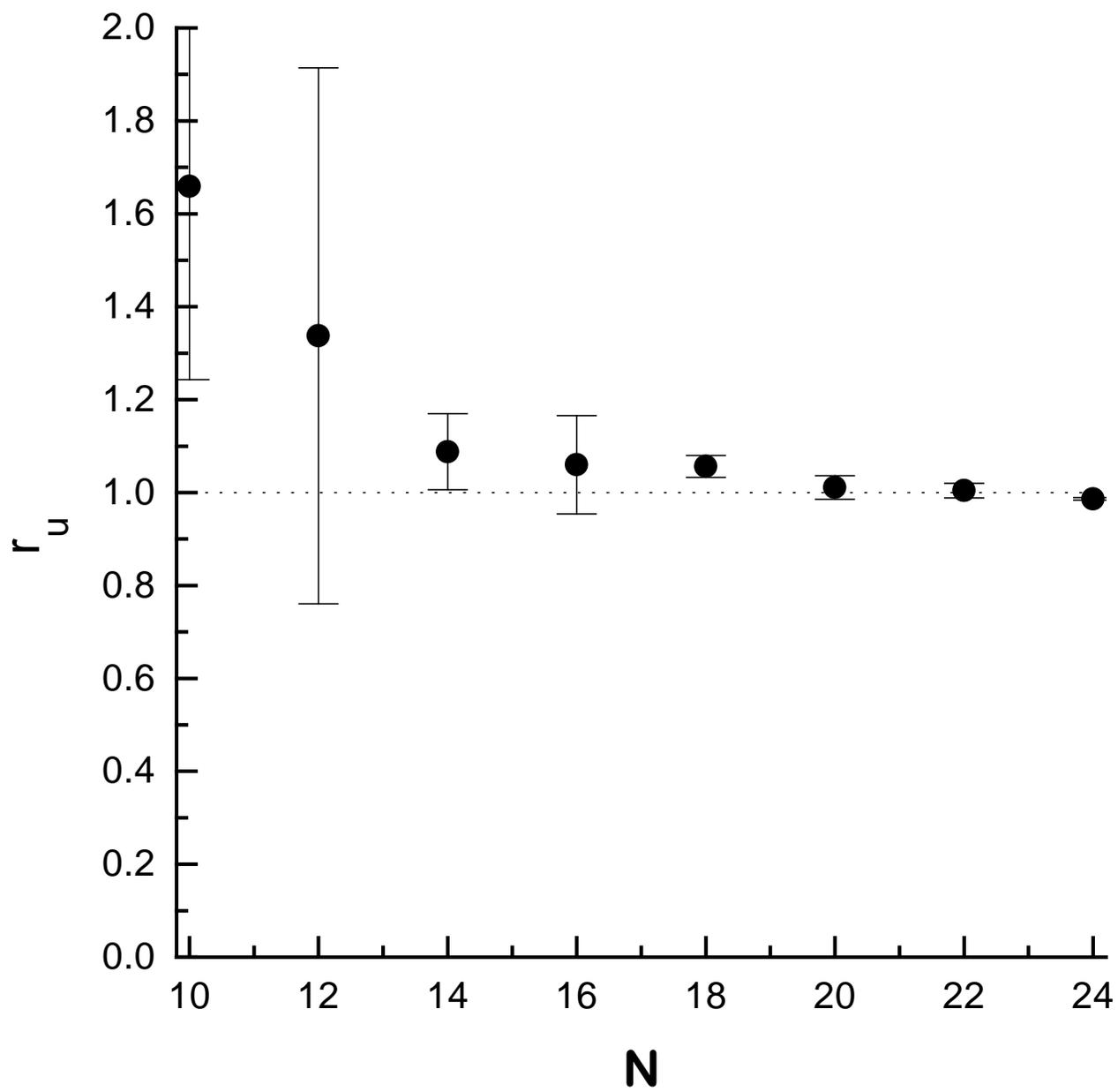
\end{document}